\documentclass[twocolumn,aps,prd,superscriptaddress,nofootinbib]{revtex4-2}
\usepackage{graphicx}
\usepackage{subcaption}
\usepackage{color}
\graphicspath{{figures/}{fig/}}
\usepackage{amsmath}
\usepackage{amssymb}
\usepackage{bm}
\usepackage{slashed}
\usepackage{epsfig}
\usepackage{amsfonts}
\usepackage{epstopdf}
\usepackage{hyperref}
\usepackage{bbm}
\usepackage{textcomp}
\usepackage{color}
\usepackage{booktabs}

\newcommand{\sect}[1]{\section{#1}}

\begin{document}
\title{ Reduction of $\epsilon$-expanded Feynman integrals}

\author{Yan-Qing Ma}
\email{yqma@pku.edu.cn}
\affiliation{School of Physics, Peking University, Beijing 100871, China}
\affiliation{Center for High Energy Physics, Peking University, Beijing 100871, China}

\author{Cong-Hao Qin}
\email{qinconghao@sjtu.edu.cn}
\affiliation{School of Physics and
Astronomy, Shanghai Jiao Tong Univeristy, Shanghai 200240, China}

\author{Ao Tan}
\email{tanao@stu.pku.edu.cn}
\affiliation{School of Physics, Peking University, Beijing 100871, China}

\author{Kai Yan}
\email{yan.kai@sjtu.edu.cn}
\affiliation{School of Physics and
Astronomy, Shanghai Jiao Tong Univeristy, Shanghai 200240, China}
\affiliation{Key Laboratory for Particle Astrophysics and Cosmology (MOE), Shanghai 200240, China}

\date{\today}

\begin{abstract}

Since Feynman integrals (FIs) at higher spacetime dimensions are free of infrared and collinear divergences—and their ultraviolet divergences can be systematically subtracted—this allows us to construct a wide range of locally finite Feynman integrals. Especially, we propose a method named $\bar{R}$-operation to subtract out ultraviolet divergences that at the same time preserves infrared and collinear safety of the original FI. By expressing these locally finite FIs in terms of master integrals and imposing constraints on their $\epsilon$-expanded forms, we reduce the $\epsilon$-expanded master integrals to a minimal basis. We provide an automated package to identify such constraints, offering a tool useful for high-order perturbative computations.

\end{abstract}

\maketitle
\allowdisplaybreaks

\sect{Introduction}

Feynman integrals (FIs) are central to quantum field theory, forming the backbone of most scattering amplitudes. A thorough understanding of FIs is therefore essential. A key property of FIs is that, within a given family, they span a finite-dimensional linear space \cite{Smirnov:2010hn}, with the basis of which are called master integrals (MIs). Using techniques such as integration-by-parts (IBP) identities \cite{Tkachov:1981wb, Chetyrkin:1981qh, Laporta:2000dsw, Gluza:2010ws, Schabinger:2011dz, vonManteuffel:2012np, Lee:2013mka, vonManteuffel:2014ixa, Larsen:2015ped, Peraro:2016wsq, Mastrolia:2018uzb, Liu:2018dmc, Guan:2019bcx, Klappert:2019emp, Peraro:2019svx, Frellesvig:2019kgj, Wang:2019mnn, Smirnov:2019qkx, Klappert:2020nbg, Boehm:2020ijp, Heller:2021qkz, Bendle:2021ueg}, any FI can be systematically reduced to a combination of MIs.

Although MIs form a minimal basis for the general spacetime dimension \( D = 4 - 2\epsilon \) in dimensional regularization \cite{Bollini:1972ui, tHooft:1972tcz}, they are not linearly independent when considering only a few terms in the \( \epsilon \)-expansion. A well-known example is the relation between 5-point and 4-point integrals at the one-loop level \cite{Bern:1993kr}:
\begin{align} \label{eq:E0}
    E_0=\sum_i a_i D_{0i} + {\cal O}(\epsilon),
\end{align}
which implies that the number of basis elements can be reduced if the \( {\cal O}(\epsilon) \) contributions are negligible. This relation also greatly simplifies one-loop computations, as \( E_0 \) is much more complex than the \( D_0 \)'s. However, even after decades, a comprehensive approach to extend the crucial relation given in Eq.~\eqref{eq:E0} to multiloop levels is still lacking. 

In practical applications, obtaining finite cross-sections or event-shape observables \cite{Henn:2019gkr, Yan:2022cye, Chicherin:2024ifn} typically requires only the finite values of FIs. 
Furthermore, for computing renormalization group equations \cite{Wilson:1973jj}, such as UV renormalization constants and DGLAP evolution kernels \cite{Gribov:1972ri, Altarelli:1977zs, Dokshitzer:1977sg}, as well as the soft anomalous dimension \cite{Henn:2023pqn, Bruser:2020bsh}, FIs up to terms divergent as $1/\epsilon$ are usually sufficient. Calculating FIs to fixed orders in the $\epsilon$-expansion can be more efficient than computing them for a general spacetime dimension. Significant efforts have been made in this direction to optimize the evaluation of FIs \cite{Gluza:2010ws, Caron-Huot:2014lda, vonManteuffel:2014qoa, Dubovyk:2022frj, Hidding:2022ycg, Hidding:2020ytt, vonManteuffel:2012np, Maierhofer:2017gsa, Klappert:2020nbg, Armadillo:2022ugh, Borowka:2018goh}. A much more powerful simplification is achievable if we can establish systematic relations, similar to Eq.~\eqref{eq:E0}, that link complex FIs to simpler counterparts up to the desired expansion order.
The relation~\eqref{eq:E0} can be interpreted as indicating that a FI, which is a linear combination of 5-point and 4-point integrals, is evanescent, ${\cal O}(\epsilon)$. The key to obtaining such evanescent or finite FIs is to construct integrals that are free from infrared (IR), collinear (CO), and ultraviolet (UV) divergences. 

A straightforward approach to avoid IR and CO divergences is to increase the spacetime dimension of the FIs, since the integration measurements can always suppress the IR and CO divergences originating from the denominators if the dimension is sufficiently high. Besides, if the rank of the numerator is not too high, UV divergences do not arise either. This allows for the construction of numerous finite FIs in higher dimensions. By using the dimension recurrence relations (DRR) \cite{Tarasov:1996br}, we can obtain linear combinations of FIs at $D=4-2\epsilon$ that are free of divergences.
It is also worth mentioning that another method of finding finite and evanescent FIs  was proposed in \cite{Gambuti:2023eqh} and \cite{delaCruz:2024xsm} by constructing special numerators directly around $D=4$. Finite FIs obtained in the above methods are unfortunately limited, because the the rank of the numerator cannot be too large.

Nevertheless, if simpler FIs that do not belong to the original integral family are allowed to be introduced, a significantly larger set of finite FIs can be constructed. The reasoning is that, for any given FI in the original family, in principle we can construct local IR, CO, and UV subtraction terms to make the rest of integral finite. Although these subtraction terms do not belong to the original family, they may be easier to handle. For UV divergences, some methods exist to perform the subtraction, such as the $R$-operation \cite{Chetyrkin:1982nn, Chetyrkin:1984xa, Herzog:2017bjx}. 
In contrast, local IR and CO subtraction terms generally lead to FIs with denominators that are linear in the loop momenta, which may not necessarily result in simpler FIs.

Prior to the development of satisfactory IR and CO subtraction terms, this paper proposes a mixed prescription to systematically derive relations like Eq.~\eqref{eq:E0} at the multiloop level. Beginning from any FI, the approach first increases the spacetime dimension to a certain value where the FI  becomes IR- and CO-finite. We then propose a modified version of UV subtraction method, named $\bar{R}$-operation, where the subtraction terms do not introduce new IR or CO divergences. Based on the finite FIs obtained in higher dimensions, we subsequently apply DRR to transform them into 4-dimensional integrals. Finally, we can generate relations for MIs after expanding the regulator, resulting in a minimal basis.

The rest of the paper is organized as follows. In Sec.~\ref{sect:reginoAnalysis}, we review the method for analyzing the regions of FIs and discuss how to constrain the MIs after expanding the regulator $\epsilon$.  In Sec. \ref{sect:eliminateDiv}, we demonstrate how to eliminate divergent regions in high dimensions, including IR, CO, and UV divergences and how to apply DRR. In Sec. \ref{sect:example}, we apply our method to several examples, ranging from one-loop to three-loop levels. In Sect. \ref{sect:package}, we introduce our package for generating relations of MIs. We conclude in Sec. \ref{sect:summary}.

\sect{General strategy}
\label{sect:reginoAnalysis}

Eq.~\eqref{eq:E0} is a specific example of relations among FIs taking the following form:
\begin{align}\label{eq:linearCB}
\sum_i c_i I_i ={\cal O}(\epsilon^n
), \quad n = 0, 1 .
\end{align}
where the left-hand side combines into finite (or evanescent) integrals in $D=4$. By systematically constructing linear combinations of FIs 
which start at the order of ${\cal O}(\epsilon^n),\,  n=0$ or $1$, we can identify crucial relations like Eq.~\eqref{eq:E0}, which relates the divergent (or ${\cal O}(\epsilon^0)-$finite) part of FIs in terms of those of simpler FIs. 
As mentioned above, we will construct IR and CO finite integrals in high spacetime dimensions, and thus we only need to focus on the case of $n=0$, which means the left hand side of Eq. \eqref{eq:linearCB} is finite. The reason is that as the dimensionality increases, a greater amount of external momenta is needed to form evanescent integrals. Consequently, the number of evanescent integrals is significantly reduced in higher dimensions.

To construct linear combinations satisfying Eq.~\eqref{eq:linearCB}, we employ the strategy of regions \cite{Beneke:1997zp}. Within this framework, any given FIs can be decomposed into distinct integration regions, which comprises singular (including IR, CO, and UV) and finite region. Singular regions are what we focus on
, which could lead to $1/\epsilon$-divergences. 
The IR and CO regions can be eliminated easily by increasing the spacetime dimension of the integral. For a given integrand, there always exists a sufficiently high dimension in which IR and CO divergences vanish, as dictated by the power-counting of the integration measurements which are proportional to the dimension. After IR and CO divergences are removed, we can focus exclusively on UV regions. In this work we propose a $\bar R$-operation method to remove the UV divergences.   
The method (introduced in Sect. \ref{sect:eliminateDiv}) performs UV subtraction without introducing any IR or CO divergences in the subtraction terms.

Once Eq.~\eqref{eq:linearCB} is obtained, one can apply IBP on it to derive a relation among MIs:
\begin{align}\label{eq:linearCB3}
    \sum_i c'_i M_i^{(D)} = {\cal O}(\epsilon^n) \quad (n=0,1),
\end{align}
where $M_i^{(D)}$ denotes the MIs in $D$ dimensions. As long as the dimension $D$ is close to an even number greater than $4$, we can subsequently transform the MIs into $4-2\epsilon$ dimensions using DRR, resulting in:
\begin{align}\label{eq:linearCB3}
    \sum_i c''_i M_i^{(4-2\epsilon)} = {\cal O}(\epsilon^n) \quad (n=0,1).
\end{align}
Expanding the integrals in the above equation in terms of $\epsilon$, we obtain a series of constraints:
\begin{align}\label{eq:linearCB4}
    \sum_i (c''_i M_i^{(4-2\epsilon)})_j = 0 \quad (j \leqslant n-1, \ n=0,1),
\end{align}
where the subscript $j$ indicates the coefficient of $\epsilon^j$. Denoting the expansion of MIs as
\begin{align}
    M_i=\sum_j M_{i,j}\, \epsilon^j,
\end{align}
the constraints obtained above  typically allow us to reduce the divergent parts of the MIs, $M_{i,j}(j\leqslant0)$, to a smaller set of coefficients, denoted as master coefficients (MCs).

By increasing the spacetime dimension to avoid IR and CO divergences and introducing UV subtractions, we can construct a corresponding locally finite FI starting from any FI around four dimensions. This procedure allows us to derive a sufficient number of relations in the form of Eq.~\eqref{eq:linearCB}.  As we will show in Sect.~\ref{sect:example}, these constraints lead to significantly fewer MCs than the number of $M_{i,j}$ terms.

\sect{Eliminating divergent regions}
\label{sect:eliminateDiv}

In this section, we discuss how to eliminate the IR and CO regions, subtract UV divergences, and to construct DRR.

\subsection{IR and CO regions}
\label{sect:IRandCO}

Although IR and CO divergences are suppressed in higher dimensions, it remains crucial to identify whether a given FI remains finite  in all IR and CO regions. The reason is that an IR- and CO-finite integral in the minimal dimension typically provides stronger constraints.

In Ref. \cite{Gambuti:2023eqh}, the Landau equations \cite{Bjorken:1959fd, Landau:1959fi, 10.1143/PTP.22.128} have been applied to find out the IR and CO regions. Here we adopt an alternative approach, by introducing a small mass term $\eta$ to each propagator, which enables us to leverage existing package {\tt asy2} \cite{Jantzen:2012mw}.

Firstly, it is useful to add a small mass term $\eta$ to each propagator appearing in the denominator, transforming the original integrand $\mathcal{N}/(D_1^{\nu_1}\cdots D_N^{\nu_N})$ into
\begin{align}
    I(\eta) = \frac{\mathcal{N}}{(D_1-\eta)^{\nu_1}\cdots (D_N-\eta)^{\nu_N}}.
\end{align}
Our task is then to analyze the integration regions of $I(\eta)$ in the limit $\eta\rightarrow 0$.

Once we expand  asymptotically as $\eta \rightarrow 0$ through expansion by regions \cite{Beneke:1997zp}, IR and CO divergences will manifest as terms in the form of $\eta^{-\kappa} \ln^i\eta$, 
since $\eta$ serves as an IR and CO regulator. 
The non-analytic regions of $\eta$ then map directly onto the IR and CO regions, which can be efficiently identified using package {\tt asy2} \cite{Jantzen:2012mw}. 
As an example, consider a one-loop integral
\begin{align}
    I(\eta) = \frac{1}{(l^2-\eta)((l-q)^2-\eta)}.
\end{align}
$I(\eta)$ has two regions, $l \sim \eta^{\frac{1}{2}}$ and $l-q \sim \eta^{\frac{1}{2}}$, corresponding to two IR regions of $I(0)$, $l = 0$ and $l-q = 0$, respectively.

The next step is to examine the power-counting behaviours of the $D-$dimensional integrals in a given region. The substitution rules for power-counting within an IR or CO region are as follows \cite{Anastasiou:2018rib, Sterman:1978bi, Libby:1978qf, Libby:1978bx}: 
\begin{align}\label{eq:rule}
    \begin{aligned}
        &\mathrm{IR}: \quad q^\mu_k \rightarrow \lambda \sigma^\mu_k , \\
        &\mathrm{CO}: \quad q^\mu_k \rightarrow x_k p_i^\mu + \lambda^2 p_j^\mu + \lambda q^\mu_{k\perp} ,
    \end{aligned}
\end{align}
where $q_k$ is the momentum of $D_k$ and $\lambda$ is a small parameter for power-counting. These rules correspond to the conditions $q_k=0$ and $q_k\parallel p_i$, respectively. Applying these substitution rules, it becomes evident that a $D$-dimensional integration measurement $\mathrm{d}^D l$ contributes a $D$ in power-counting, which explains why IR and CO divergences are suppressed in high dimensions.

\subsection{UV regions}
\label{subsect:uvregion}

The power-counting theorem \cite{Dyson:1949ha, Weinberg:1959nj} states that a FI is UV finite if the integral itself and its subintegrations have  negative superficial degrees of divergence.
%
In order to analyze all UV regions, including the subintegration regions, it is helpful introduce the following notation for the linear combinations of loop momenta $l_i$'s appearing in the propagators: 
\begin{align}\label{eq:LoopComb}
   {\{ \sum_{j=1}^L e_{i j} l_j \ | \ D_i=(\sum_{j=1}^{L} e_{i j} l_j+r_i)^2 - m^2_j \}}, 
\end{align}
where $L$ is the number of loops, $e_{i j}$'s are typically $0$ or $\pm 1$, and $r_i$ is the linear combination of external momenta for $i$-th propagator $D_i$. A UV region is characterized by the scaling behavior of these combinations of $l_i$'s, namely, some remain finite while others go to infinity. 
For instance, for a 2-loop integral, there are only three linear combinations of $l_i$'s, $\{ l_1, l_2, l_1 + l_2 \}$, leading to 4 possible UV regions, 
\begin{align}
    \begin{aligned}
        &\{ l_2=S, l_1 \rightarrow \infty \}, \\
        &\{ l_1=S, l_2 \rightarrow \infty \}, \\
        &\{ l_1+l_2=S, l_1 \rightarrow \infty \}, \\
        &\{ l_1,l_2 \rightarrow \infty \},
    \end{aligned}
\end{align}
where $S$ denotes a small momentum.

Once all UV regions are identified, the next step is to apply UV subtractions. In the $R$-operation method \cite{Chetyrkin:1982nn, Chetyrkin:1984xa, Herzog:2017bjx}, a subtraction formula proposed: 
\begin{align}\label{eq:LLUVsub}
    G_{\text{UVCT}} = \sum_{k=1}^{L}\sum_{r_1 \prec r_2 \prec \cdots \prec r_k  \atop r_i \in \mathcal{R}}(-1)^{k}G_{r_1 r_2 \cdots r_k},
\end{align}
which satisfies that $G+G_{\text{UVCT}}$ is UV-finite. 
The notations in this formula are defined as follows: $\mathcal{R}$ is the set of UV regions, $r_1 \prec r_2$ means $r_1$ is a sub-region of $r_2$, and $G_{r_1 r_2 \cdots r_k}\equiv (G_{r_1 r_2 \cdots r_{k-1}})_{r_k}$ is the UV counterterm of $G_{r_1 r_2 \cdots r_{k-1}}$ in the UV region $r_k$.
To check whether $r_1 \prec r_2$, we need to list all the elements of the set~\eqref{eq:LoopComb} that are small in region $r_1$, denoting this subset as $\mathcal{S}_1$. Similarly, let $\mathcal{S}_2$ be the corresponding subset for region $r_2$. Then $r_1 \prec r_2$ if and only if $\mathcal{S}_1 \supsetneq \mathcal{S}_2$.

However, the counterterm $G_r$ naively introduced in $R$-operation \cite{Chetyrkin:1982nn, Chetyrkin:1984xa, Herzog:2017bjx} can not be directly used for our purpose. The issue arises because the counterterms of local UV subtraction may introduce IR and CO divergences, which were initially suppressed by the numerator $\mathcal{N}$. To solve this problem, we will introduce a modified version, $\bar{R}$-operation. In our approach, the subtraction formula is the same as ${R}$-operation in Eq.~\eqref{eq:LLUVsub}, but the UV counterterms can preserve IR- and CO-safety.

\subsection{Subtraction terms in $\bar{R}$-operation}
\label{sec:rbar}

Now let us explain how to construct UV counterterms in $\bar{R}$-operation that can preserve IR- and CO-safety. We will take $G_r$ as an example to show this, because the counterterm $G_{r_1 r_2 \cdots r_k} = (G_{r_1 r_2 \cdots r_{k-1}})_{r_k}$ can be constructed in the same way. To avoid the appearance of IR and CO divergences in $G_r$, two key points must be considered. First, an additional mass term should be introduced into the propagators of $G_r$ to prevent it from becoming a scaleless integral, which usually has IR divergences. Second, the structure of the numerator should be preserved as much as possible, as this helps to cancel the IR and CO divergences. 

To implement the first point, we use the following variable substitution: 
\begin{align}\label{eq:identitysubstitute}
    \begin{cases}
        l_i \cdot l_j = \hat{l}_i \cdot \hat{l}_j + c_{i j} m^2 \\
        p_i \cdot l_j = p_i \cdot\hat{l}_j
    \end{cases},
\end{align}
where $m^2$ is the newly introduced mass term in the propagators of $G_r$. It can be understood as only substituting components of loop momenta in the space perpendicular to external momenta. The freedom of the substitution enables us to choose coefficient $c_{i i}$'s as $1$. In the special case of 2-loop situation, $c_{i j}$'s can be choses to satisfy that $\hat{l}_1^2 = l_1^2-m^2$, $\hat{l}_2^2 = l_2^2-m^2$ and $(\hat{l}_1 + \hat{l}_2)^2 = (l_1+l_2)^2-m^2 $. 

For the second point, we only apply this substitution to the denominator while the variables in the numerator remain unchanged.

Suppose that in region $r$, the loop momenta that go to infinity are denoted as $L_i$'s and the linear independent set of the small combinations of loop momenta are denoted as $S_i$'s. Denoting $\hat{L}_i$'s and $\hat{S}_i$'s as the corresponding variables after the substitution in Eq.~\eqref{eq:identitysubstitute}, we can rewrite $G$ as a function of these momenta: 
\begin{align}\label{eq:substitutedG}
    G(L_i, S_i, \hat{L}_i, \hat{S}_i) = \frac{\mathcal{N}(L_i, S_i)}{D_1^{\nu_1}(\hat{L}_i, \hat{S}_i) \cdots D_N^{\nu_N}(\hat{L}_i, \hat{S}_i)}, 
\end{align}
where the other variables, like external momenta and masses, are omitted for brevity.

With the condition that $L_i$'s and $\hat{L}_i$'s tend to infinity, we expand $G$ in \eqref{eq:substitutedG} as: 
\begin{align}\label{eq:expandedG}
    G(\Lambda L_i, S_i, \Lambda\hat{L}_i, \hat{S}_i) = \sum_{k=-\infty}^{\infty} \frac{g_k(L_i, S_i, \hat{L}_i, \hat{S}_i)}{\Lambda^k},
\end{align}
where $\Lambda$ is a large parameter, and in fact, the summation starts from a finite $k$. With this expansion, the counterterm $G_r$ is defined as follow:
\begin{align}\label{eq:counterterm}
    G_r=\sum_{k=-\infty}^{w D}g_k(L_i, S_i, \hat{L}_i, \hat{S}_i),
\end{align}
where $w$ is the number of $L_i$'s, $D$ is the dimension of the integral and $w D$ is the superficial degree of the integration measurements in region $r$. The expansion up to this order is sufficient to fully subtract the UV divergences in region $r$.

Next we will prove that the counterterms $G_r$ do not introduce new IR and CO divergences. Given the extra mass in the propagators for $L_i$'s [via the substitution~\eqref{eq:identitysubstitute}], only IR and CO regions of the momenta $S_i$'s can emerge in the UV counterterm $G_r$. Suppose $R$ is one of these IR or CO regions, then it suffices to show that $G_r$ remains finite in the region $R$. Since IR and CO regions of $G_r$  must inherit from those of $G$, $R$ is also an IR or CO region of $G$. Meanwhile, $G$ remains finite in $R$ by construction. Therefore, the numerator of Eq.~\eqref{eq:substitutedG}, behaving as  $\mathcal{N} = \mathcal{O}(\lambda^a)$ under rules~\eqref{eq:rule} of the region $R$,  will suppress the negative powers of $\lambda$ introduced by the denominator.

On the one hand, according to the definition, the negative powers $\lambda$ introduced by the denominator of $G_r$ are the same as that of $G$, because the expansion in Eq.~\eqref{eq:expandedG} only applies for large denominators in $R$. 
On the other hand, we express the numerator $\mathcal{N}$ as a sum of homogeneous function:
\begin{align}
    \mathcal{N}(L_i, S_i) = \sum_k \varphi_k(L_i, S_i),
\end{align}
where $\varphi_k$ is a $k$-degree homogeneous function of $L_i$'s. 
Then each $\varphi_k$ must independently satisfy $\varphi_k = \mathcal{O}(\lambda^a)$ because there can not be cancellation between $\varphi_k$'s.
Now we turn to Eq.~\eqref{eq:substitutedG} and find that
\begin{align}\label{eq:LambdaG}
    G(\Lambda L_i, S_i, \Lambda\hat{L}_i, \hat{S}_i) = \frac{\sum_k \Lambda^k \varphi_k(L_i, S_i)}{D_1^{\nu_1}(\Lambda\hat{L}_i, \hat{S}_i) \cdots D_N^{\nu_N}(\Lambda\hat{L}_i, \hat{S}_i)}, 
\end{align}
Hence each term $g_k$ in Eq.~\eqref{eq:expandedG} is proportional to one of $\varphi_k$'s. This property ensures that $g_k$'s are still finite in the region $R$, and thus $G_r$ is finite in $R$.

The above derivation is also applicable to $G_{r_1 r_2 \cdots r_k}$, resulting that if $G$ is finite, $G+G_{\text{UVCT}}$ is finite as well.

We provide an example in Appendix \ref{appendix:UV}, to demonstrate how to compute the UV counterterms of the two-loop sunrise integrand which is IR- and CO-finite.

\subsection{Dimensional recurrence relations}

There are two types of DRR \cite{Tarasov:1996br,Lee:2009dh}, raising and lowering relations. It is useful to introduce the operators $n^{\pm}$, acting on the FIs as follows: 
\begin{align}
    &\begin{aligned}
        &\mathbf{i}^+ I^{(D)}(n_1, \cdots, n_i , \cdots, n_N) \\
        =& n_i I^{(D)}(n_1, \cdots, n_i + 1 , \cdots, n_N),
    \end{aligned} \\
    &\begin{aligned}
        &\mathbf{i}^- I^{(D)}(n_1, \cdots, n_i , \cdots, n_N) \\
        =& I^{(D)}(n_1, \cdots, n_i - 1 , \cdots, n_N).
    \end{aligned}
\end{align}
Then the raising and lowering relations are:
\begin{align}
    \label{eq:raisingDRR}
    &I^{(D-2)} = (-1)^L \mathcal{U}(\mathbf{1}^{+}, \cdots, \mathbf{K}^{+})I^{(D)}, \\
    \label{eq:loweringDRR}
    &I^{(D+2)} = \frac{(-2)^L (V(p_1, \cdots, p_E))^{-1} }{(D - E - L + 1)_L}P(\mathbf{1}^{-}, \cdots, \mathbf{N}^{-})I^{(D)},
\end{align}
where $\mathcal{U}$ is the Symanzik polynomial in the Feynman parameter representation, $K \leqslant N$ is the number of propagators in the denominator, $V(q_1, \cdots, q_n) = \det(q_i \cdot q_j)$, $P(D_1, \cdots, D_N) = V(l_1, \cdots, l_L, p_1, \cdots p_E)$ and $(x)_n = x(x+1)\cdots(x+n-1)$.

Along with IBP, the DRR can construct the relations of MIs between different dimensions, which help converting the MIs defined in any even dimensions into 4 dimensions. To obtain the relations, it is much more efficient to use the Eq.~\eqref{eq:raisingDRR} rather than~\eqref{eq:loweringDRR}, since the former only increases the dots by $L$ while the latter increases the rank by $2L$.

\sect{Examples}
\label{sect:example}

In this section, we present some explicit examples to illustrate  how we can find relations of MIs in the $\epsilon$-expansion form. First, we consider the simple one-loop pentagon to show how to reproduce the relation Eq.~\eqref{eq:E0}. Then, we apply our method to the 2-loop double box and double pentagon to explore how much constraints we can impose on the MIs. Finally, we consider a 3-loop example to validate the correctness of our method in higher-loop situation.

\subsection{One-loop pentagon}
\begin{figure}[htbp]
    \centering
    \includegraphics[width=0.12\textwidth]{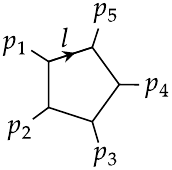}
    \caption{One-loop pentagon.}
    \label{fig:pentagon}
\end{figure}

The propagators of one-loop pentagon are defined as $D_1=l^2$, $D_2=(l-p_1)^2$, $D_3=(l-p_1-p_2)^2$, $D_4=(l-p_1-p_2-p_3)^2$, $D_5=(l-p_1-p_2-p_3-p_4)^2$, with $p_i^2=0$, $p_1+p_2+p_3+p_4+p_5=0$ and $(p_i+p_j)^2=s_{ij}$. 

By introducing a small parameter $\eta$ to each propagator and use {\tt asy2} to compute regions, we get five nontrivial regions, $\{0,0,1,1,1\}$, $\{0,-1,-1,0,0\}$, $\{0,0,-1,-1,0\}$, $\{0,0,0,-1,-1\}$ and $\{0,1,1,1,0\}$, which correspond to five collinear regions, $l \parallel p_1$, $l-p_1 \parallel p_2$, $l-p_1-p_2 \parallel p_3$, $l-p_1-p_2-p_3 \parallel p_4$ and $l \parallel p_5$.

With these IR and CO regions and power-counting substitution rules~\eqref{eq:rule}, we can find that the 6-dimensional master integral $I^{(6-2\epsilon)}(1,1,1,1,1)$ is IR- and CO-finite, and also UV-finite. The dimension difference equation of it is given by:
\begin{align}
    \begin{aligned}
        I^{(6-2\epsilon)}(1,1,1,1,1) &\propto \frac{c_1}{\epsilon} M_1 + \frac{c_2}{\epsilon} M_2 \\
        &+ \frac{c_3}{\epsilon} M_3 + \frac{c_4}{\epsilon} M_4 \\
        &+ \frac{c_5}{\epsilon} M_5 - \frac{1}{\epsilon} I(1,1,1,1,1),
    \end{aligned} 
\end{align}
where, for simplicity, an integral is in $4-2\epsilon$ dimensions if there is no superscript. In the above relation, $M_1 = I(0,1,1,1,1), M_2 = I(1,0,1,1,1), \ldots, M_5 = I(1,1,1,1,0)$, 
and
\begin{align}\label{eq:pentadrr}
    \begin{aligned}
        c_1 &= \frac{-s_{12}s_{23}+s_{23}s_{34}-s_{34}s_{45}+s_{45}s_{51}+s_{51}s_{12}}{2s_{12}s_{45}s_{51}}, \\
        c_2 &= \frac{s_{12}s_{23}-s_{23}s_{34}+s_{34}s_{45}-s_{45}s_{51}+s_{51}s_{12}}{2s_{12}s_{23}s_{51}}, \\
        c_3 &= \frac{s_{12}s_{23}+s_{23}s_{34}-s_{34}s_{45}+s_{45}s_{51}-s_{51}s_{12}}{2s_{12}s_{23}s_{34}}, \\
        c_4 &= \frac{-s_{12}s_{23}+s_{23}s_{34}+s_{34}s_{45}-s_{45}s_{51}+s_{51}s_{12}}{2s_{23}s_{34}s_{45}}, \\
        c_5 &= \frac{s_{12}s_{23}-s_{23}s_{34}+s_{34}s_{45}+s_{45}s_{51}-s_{51}s_{12}}{2s_{34}s_{45}s_{51}}.
    \end{aligned}
\end{align}

From Eq.~\eqref{eq:pentadrr}, we cen deduce that
\begin{align}
    \begin{aligned}
        I(1,1,1,1,1)_0 &= c_1 M_{1,0} + c_2 M_{2,0} \\
        &+ c_3 M_{3,0} + c_4 M_{4,0} \\
        &+ c_5 M_{5,0},
    \end{aligned} 
\end{align}
where the subscript $0$ means the coefficient of $\epsilon^0$.

Moreover, we construct a few more finite FIs beginning from IR- and CO-finite  integrals $I^{(6-2\epsilon)}(n_1,\cdots,n_5)$, where some of $n_i$'s can be negative, and applying UV subtraction. The constraints imposed by these finite FIs result in that divergent parts of all MIs in this family can be expressed in terms of five 2-point integrals and one-loop vacuum integral: $I(0,0,1,0,1)_0$, $I(0,1,0,0,1)_0$, $I(0,1,0,1,0)_0$, $I(1,0,0,1,0)_0$, $I(1,0,1,0,0)_0$ and $I_{UV,-1}$, where the vacuum integral is defined as:
\begin{align}
    I^{(D)}_{UV} = \int\frac{\mathrm{d}^D l}{(2\pi)^{D}}\frac{1}{l^2-1}.
\end{align}
For example:
\begin{align}\label{eq:pentaRela}
    &I(1,1,1,1,1)_{-2} = \frac{\sum_{circle} s_{12}s_{23}}{s_{12}s_{23}s_{34}s_{45}s_{51}}I_{UV,-1}.
\end{align}

It is not surprising to reach such a result, because functions of divergent parts at one-loop level are no more than $\ln(-s_{12})$, $\ln(-s_{23})$, $\ln(-s_{34})$, $\ln(-s_{45})$, $\ln(-s_{51})$ and constants.

\subsection{Two-loop double box}
\begin{figure}[htbp]
    \centering
    \includegraphics[width=0.18\textwidth]{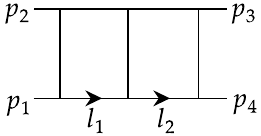}
    \caption{Two-loop double box.}
    \label{fig:doublebox}
\end{figure}

The propagators of double box are $l_1^2$, $(l_1-p_1)^2$, $(l_1-p_1-p_2)^2$, $l_2^2$, $(l_2-p_1-p_2-p_3)^2$, $(l_2-p_1-p_2)^2$, $(l_1-l_2)^2$ and the irreducible scalar products (ISPs) are selected as $(l_1-p_3)^2$, $(l_2-p_1)^2$, with $p_i^2=0$, $p_1+p_2+p_3+p_4=0$, $(p_1+p_2)^2=s$, and $(p_2+p_3)^2=t$. 

By applying {\tt asy2}, there are 13 soft and collinear regions to be considered: $\{ l_1-p_1 \parallel p_2,l_2 \parallel p_4 \}$, $\{ l_1-p_1 \parallel p_2,l_2+p_4 \parallel p_3 \}$, $\{ l_1-p_1 \parallel p_2,l_2-p_1 \parallel p_2 \}$, $\{ l_1-p_1 \parallel p_2 \}$, $\{ l_1-p_1-p_2 \parallel p_3,l_2+p_4 \parallel p_3 \}$, $\{ l_2 \parallel p_4 \}$, $\{ l_2+p_4 \parallel p_3 \}$, $\{ l_1-l_2 =0 \}$, $\{ l_1 \parallel p_1,l_2 \parallel p_4 \}$, $\{ l_1 \parallel p_1,l_2 \parallel p_1 \}$, $\{ l_1 \parallel p_1,l_2+p_4 \parallel p_3 \}$, $\{ l_1 \parallel p_1 \}$, $\{ l_1 \parallel p_4,l_2 \parallel p_4 \}$. These regions are consistent with the results in \cite{Gambuti:2023eqh}.

It is interesting that we find the divergent parts of the MIs in the topsector, $M_1 = I(1,1,1,1,1,1,1,0,0)$ and $M_2 = I(1,1,1,1,1,1,1,0,-1)$, can be expressed in terms of MIs from the subsectors. This result is derived from two 6-dimensional integrals, $I^{(6-2\epsilon)}(1,1,1,1,1,1,1,0,0)$ and $I^{(6-2\epsilon)}(1,1,1,1,1,1,2,0,0)$, which are IR-, CO- and UV-finite. Using IBP and DRR (see Appendix \ref{appendix:dbdrr} for details of double box), these two integrals can be represented as:
\begin{align}
    &\begin{aligned}
        \mathcal{O}(\epsilon^0)=&I^{(6-2\epsilon)}(1,1,1,1,1,1,1,0,0) \\
        =& \ \frac{s^2((-1+3\epsilon)s + (-1+2\epsilon)t)}{4(-1+2\epsilon)^3(s+t)}M_1 \\
        &- \frac{s^2((-3+9\epsilon)s + (-3+8\epsilon)t)}{4(-1+2\epsilon)^3t(s+t)}M_2 \\
        &+ (\text{subsectors}),
    \end{aligned} \\
    &\begin{aligned}
       \mathcal{O}(\epsilon^0)= &I^{(6-2\epsilon)}(1,1,1,1,1,1,2,0,0) \\
        =& -\frac{s^2}{4(-1+2\epsilon)(s+t)}M_1 \\
        &+ \frac{s(3s+2t)}{4(-1+2\epsilon)t(s+t)}M_2 \\
        &+ (\text{subsectors}).
    \end{aligned}
\end{align}
Applying Gauss elimination to the 2 equations above, we note that each step in the elimination will remain $\mathcal{O}(\epsilon^0)$ because the coefficients and the determinant of the coefficient matrix are all $\mathcal{O}(\epsilon^0)$. Finally we obtain the result:
\begin{align}
    M_i = (\text{subsectors}) + \mathcal{O}(\epsilon^0). \quad (i=1,2) 
\end{align}

Furthermore, we can do the same thing to the 6-dimensional integrals with rank up to 2 and dots up to 1. Consequently, we find that the divergent parts of all MIs can be reduced to 14 MCs: 
\begin{align}\label{eq:dbdivmaster}
    \begin{aligned}
        &I(1,1,0,0,1,1,1,0,0)_{-1}, \\
        &I(1,1,1,0,1,0,1,0,0)_{-1, 0}, \\
        &I(1,0,1,0,1,0,1,0,0)_{0, 1}, \\
        &I(1,0,1,1,0,1,0,0,0)_{1}, \\
        &I(0,1,0,0,1,0,1,0,0)_{0,1,2}, \\
        &I(1,0,0,0,0,1,1,0,0)_{1,2}, \\
        &I_{UV1,-2,-1}, I_{UV4,-1}, 
    \end{aligned}
\end{align}
where the integrals corresponding to the last 2 expansion coefficients are: 
\begin{align}
    \begin{aligned}
        &I_{UV1}^{(D)} = \int\frac{\mathrm{d}^D l_1 \mathrm{d}^D l_2}{(2\pi)^{2D}}\frac{1}{(l_1^2-1)l_2^2(l_2-p_1-p_2)^2}, \\
        &I_{UV4}^{(D)} = \int\frac{\mathrm{d}^D l_1 \mathrm{d}^D l_2}{(2\pi)^{2D}}\frac{1}{(l_1^2-1)(l_2^2-1)}.
    \end{aligned}
\end{align}
For a concrete example, $M_{1,-2}$ is equal to
\begin{align}
    \begin{aligned}
        & \frac{6}{s t}I(1,1,1,0,1,0,1,0,0)_{-1} \\
        +& \frac{11}{s^2 t^2}I(0,1,0,0,1,0,1,0,0)_{0} \\
        +& \frac{1}{s^2 t}I(1,0,1,0,1,0,1,0,0)_{0} \\
        +& \frac{2}{s^2 t^2}I(0,1,0,0,1,0,1,0,0)_{1} \\
        +& \frac{8}{s^3 t}I(1,0,0,0,0,1,1,0,0)_{1} \\
        +& \frac{153}{4s^2 t }I_{UV1,-2}+\frac{9}{s^2 t }I_{UV1,-1} \\
        -&\frac{9}{2s^2 t }I_{UV4,-1}.
    \end{aligned}
\end{align}

There are 8 MIs for double box and $4\times8=32$ expansion coefficients if we expand the MIs up to $\epsilon^{-1}$ order. Relations generated by finite FIs can reduce these 32 coefficients by half.

If we expand the MIs up to $\epsilon^{0}$ order, we find that they can be reduced to integrals \eqref{eq:dbdivmaster}, along with 3 extra coefficients: $I(1,1,0,0,1,1,1,0,0)_{0}$, $I(1,1,1,1,1,1,1,0,0)_{0}$, $I(1,1,1,1,1,1,1,0,-1)_{0}$.

\subsection{Two-loop double pentagon}
\begin{figure}[htbp]
    \centering
    \includegraphics[width=0.18\textwidth]{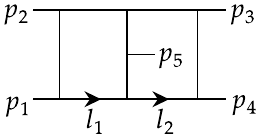}
    \caption{Two-loop double pentagon.}
    \label{fig:doublepentagon}
\end{figure}

The propagators of double pentagon are $l_1^2$, $(l_1-p_1)^2$, $(l_1-p_1-p_2)^2$, $l_2^2$, $(l_2+p_4)^2$, $(l_2+p_3+p_4)^2$, $(l_1-l_2)^2$, $(l_1-l_2-p_1-p_2-p_3-p_4)^2$  and the ISP's are selected as $(l_1-p_3)^2$, $(l_1-p_4)^2$, $(l_2-p_1)^2$, with $p_i^2=0$, $p_1+p_2+p_3+p_4+p_5=0$  and $(p_i+p_j)^2=s_{ij}$. 
There are 21 soft and collinear regions to be considered: $\{ l_1-p_1 \parallel p_2,l_2 \parallel p_4 \}$, $\{ l_1-p_1 \parallel p_2,l_2+p_4 \parallel p_3 \}$, $\{ l_1-p_1 \parallel p_2,l_2+p_3+p_4 \parallel p_2 \}$, $\{ l_1-p_1 \parallel p_2, l_1-l_2 \parallel p_5 \}$, $\{ l_1-p_1 \parallel p_2 \}$, $\{ l_1-p_1-p_2=0, l_2+p_3+p_4=0 \}$, $\{ l_1-p_1-p_2 \parallel p_3, l_2+p_4 \parallel p_3 \}$, $\{ l_1-p_1-p_2 \parallel p_5, l_2+p_3+p_4 \parallel p_5 \}$, $\{ l_2 \parallel p_4, l_1-l_2 \parallel p_5 \}$, $\{ l_2 \parallel p_4 \}$, $\{ l_2+p_4 \parallel p_3, l_1-l_2 \parallel p_5 \}$, $\{ l_2+p_4 \parallel p_3 \}$, $\{ l_1-l_2 \parallel p_5\}$, $\{ l_1 \parallel p_1,l_2 \parallel p_4 \}$, $\{ l_1 \parallel p_1,l_2 \parallel p_1 \}$, $\{ l_1 \parallel p_1,l_2+p_4 \parallel p_3 \}$, $\{ l_1 \parallel p_1, l_1-l_2 \parallel p_5 \}$, $\{ l_1 \parallel p_1 \}$, $\{ l_1=0,l_2=0 \}$, $\{ l_1 \parallel p_4,l_2 \parallel p_4 \}$, $\{ l_1 \parallel p_5,l_2 \parallel p_5 \}$.

We generate finite integral relations by using 6-dimensional integrals (up to rank 3 and dots 1) and 8-dimensional integrals (up to rank 1 and dots 2) at a numerical point. There are 108 MIs for the double pentagon. If we only focus on the divergent parts, they can be reduced to 127 MCs that are all from the subsector integrals. If we also include the $\epsilon^0$ order, they can be reduced to 230 MCs. Unlike the one-loop pentagon, the $\epsilon^0$ terms of 3 out of the 9 MIs will be retained in the MCs. Furthermore, both sets of MCs contain 4 coefficients that originate from UV counterterms. 
 
The distributions of the MCs are given in Tab.~\ref{tab:dp}, showing both the case where only the divergent parts are considered and the case where the $\epsilon^0$ terms are included.

\begin{table}[htbp]
    \centering
    \begin{tabular}{ccccccc}
        \toprule 
        $\epsilon^{-4}$ & $\epsilon^{-3}$ & $\epsilon^{-2}$ & $\epsilon^{-1}$ & $\epsilon^{0}$ & $\epsilon^{1}$ & $\epsilon^{2}$ \\
        \midrule 
        0 & 0 & 11 & 58 & 35 & 15 & 8 \\
        \bottomrule
    \end{tabular}

    \vspace{0.2cm}
    
    \begin{tabular}{cccccccc}
        \toprule 
        $\epsilon^{-4}$ & $\epsilon^{-3}$ & $\epsilon^{-2}$ & $\epsilon^{-1}$ & $\epsilon^{0}$ & $\epsilon^{1}$ & $\epsilon^{2}$ & $\epsilon^{3}$\\
        \midrule 
        0 & 0 & 11 & 58 & 99 & 40 & 14 & 8 \\
        \bottomrule
    \end{tabular}

    \caption{The number of MCs in each order of $\epsilon$ for the double pentagon family. The MIs are expanded to order $\epsilon^{-1}$ in the upper case and to order $\epsilon^{0}$ in the lower case.}
    \label{tab:dp}
\end{table}

\subsection{3-loop example}
\begin{figure}[htbp]
    \centering
    \includegraphics[width=0.14\textwidth]{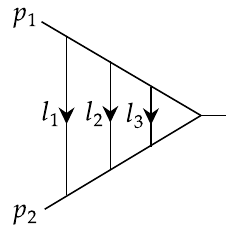}
    \caption{A simple 3-loop example.}
    \label{fig:3loop}
\end{figure}

To show the validity of our method in 3-loop situation, we study a simple example showed in the figure~\ref{fig:3loop}. The propagators are $l_1^2$, $l_2^2$, $l_3^2$, $(l_1-p_1)^2$, $(l_1+l_2-p_1)^2$, $(l_1+l_2+l_3-p_1)^2$, $(l_1+p_2)^2$, $(l_1+l_2+p_2)^2$, $(l_1+l_2+l_3+p_2)^2$ and the ISP's are selected as $l_1 \cdot l_2$, $l_1 \cdot l_3$, $l_2 \cdot l_3$, with $p_1^2=p_2^2=0$ and $(p_1+p_2)^2=1$.

When we use asy2 to analyze the regions, it gives 33 regions in total. For integrals with no dots, there are only 6 collinear regions that can cause divergences: $\{l_1 \parallel p_1\}$, $\{l_1 \parallel p_2\}$, $\{l_1,l_2 \parallel p_1\}$, $\{l_1,l_2 \parallel p_2\}$, $\{l_1,l_2,l_3 \parallel p_1\}$, $\{l_1,l_2,l_3 \parallel p_2\}$.

As for UV, there are 12 regions:
\begin{align}\label{eq:3loopUVregion}
    \begin{aligned}
        &1: \{ l_1+l_2=S_1, l_3=S_2, l_1\rightarrow\infty \}, \\
        &2: \{ l_1=S_1, l_2=S_2, l_3\rightarrow\infty \}, \\
        &3: \{ l_2=S_1, l_1+l_3=S_2, l_1\rightarrow\infty \}, \\
        &4: \{ l_2=S_1, l_3=S_2, l_1\rightarrow\infty \}, \\
        &5: \{ l_1=S_1, l_2+l_3=S_2, l_2\rightarrow\infty \}, \\
        &6: \{ l_1=S_1, l_3=S_2, l_2\rightarrow\infty \}, \\
        &7: \{ l_1+l_2+l_3=S, l_1,l_2\rightarrow\infty \}, \\
        &8: \{ l_3=S, l_1,l_2\rightarrow\infty \}, \\
        &9: \{ l_1+l_2=S, l_1,l_3\rightarrow\infty \}, \\
        &10: \{ l_2=S, l_1,l_3\rightarrow\infty \}, \\
        &11: \{ l_1=S, l_2,l_3\rightarrow\infty \}, \\
        &12: \{ l_1,l_2,l_3\rightarrow\infty \}. \\
    \end{aligned}
\end{align}

We generate some finite integrals with dimensions no higher than 6 (up to rank 2 and dots 1), then find out that divergent parts of all 9 MIs can be reduced to 10 MCs. Only one of MCs, $I(1,1,1,1,0,1,0,1,1,0,0,0)_{-1}$, originates from the original family, and the rest come from 6 families appearing in UV regions 1, 2, 5, 7, 12. We computed the numerical results of MIs and verified that the reduction relations are correct.

If we expand the MIs up to $\epsilon^{0}$ order, the number of MCs increases to 17, incorporating the additional terms: 
\begin{align*}
    \begin{aligned}
        &I(1,1,1,1,0,1,0,1,1,0,0,0)_{0}, \\
        &I(0,1,1,1,0,1,0,1,0,0,0,0)_{1}, \\
        &I(1,1,1,0,0,1,0,0,1,0,0,0)_{1}, \\
        &I(1,1,1,0,1,0,0,0,1,0,0,0)_{1}, \\
        &I(0,1,1,1,0,0,0,0,1,0,0,0)_{2} 
    \end{aligned}
\end{align*}
along with 2 coefficients from UV region 1. The distributions of the MCs  are shown in Tab.~\ref{tab:3loop}.

\begin{table}[htbp]
    \centering
    \begin{tabular}{ccccccc}
        \toprule 
        $\epsilon^{-6}$ & $\epsilon^{-5}$ & $\epsilon^{-4}$ & $\epsilon^{-3}$ & $\epsilon^{-2}$ & $\epsilon^{-1}$ & $\epsilon^{0}$ \\
        \midrule 
        0 & 0 & 0 & 0 & 2 & 6 & 2 \\
        \bottomrule
    \end{tabular}

    \vspace{0.2cm}
    \begin{tabular}{ccccccccc}
        \toprule 
        $\epsilon^{-6}$ & $\epsilon^{-5}$ & $\epsilon^{-4}$ & $\epsilon^{-3}$ & $\epsilon^{-2}$ & $\epsilon^{-1}$ & $\epsilon^{0}$ & $\epsilon^{1}$ & $\epsilon^{2}$\\
        \midrule 
        0 & 0 & 0 & 0 & 2 & 6 & 4 & 4 & 1 \\
        \bottomrule
    \end{tabular}

     \caption{The number of MCs in each order of $\epsilon$ for the 3-loop example. The MIs are expanded to order $\epsilon^{-1}$ in the upper case and to order $\epsilon^{0}$ in the lower case.}
     \label{tab:3loop}
\end{table}

\subsection{4-dimensional approach}
Besides constructing finite FIs in high dimensions, an alternative approach  is to construct finite and evanescent FIs directly in 4 dimensions. In this approach, we first use the method given in Refs.~\cite{Gambuti:2023eqh, delaCruz:2024xsm} to construct a numerator $\mathcal{N}$ ensuring that the integration of $\mathcal{N}/(D_1^{\nu_1}\cdots D_N^{\nu_N})$ is IR- and CO-finite. Then, using our proposed $\bar{R}$-operation to subtract out UV divergences, introduced in Sect. \ref{sec:rbar}, a totally locally finite FI can be obtained. Note that, in this approach, it is easy to construct FIs with no dots, but it becomes much harder if we want nonzero dots.

We also apply this approach on all the examples shown in this section, for comparison with the high-dimension approach. We generate a lot of locally finite FIs without dots in 4 dimensions and find that the MCs are consistent with those obtained through high-dimension approach.

In fact, it is somewhat surprised to obtain such a result. On the one hand, according to Eq.~\eqref{eq:loweringDRR}, each locally IR- and CO-finite integral in high even dimensions corresponds to a locally IR- and CO-finite integral in 4 dimensions,  with the same number of dots. This implies that the relations provided by the former cannot be more than those provided by the latter, if integrals with dots are included. On the other hand, the integrals we generate in 4 dimensions have no dots while those generated in high-dimension approach have dots. 

\sect{Brief introduction of the package}
\label{sect:package}

Our package \textbf{FFI} is written in \textsc{Mathematica} and can be used to discover the relations of MIs in the $\epsilon$-expansion. It utilizes the \textbf{asy2} \cite{Jantzen:2012mw} to analyze IR and CO regions, \textbf{Singular} \cite{DGPS} for ideal computation, \textbf{Blade} \cite{Guan:2024byi} for IBP reduction, and \textbf{FiniteFlow} \cite{Peraro:2019svx} to solve linear equations. The full package can be downloaded from 
\begin{align*}
    \text{\url{https://github.com/Tanao-pku/FFI}}
\end{align*}
To use \textbf{FFI}, the first step is to use \texttt{DefineFamily} to define a family object: 
\begin{align*}
    \begin{aligned}
        &\text{\texttt{DefineFamily[ Family, Propagators, ISPs,}} \\
        &\text{\texttt{    PropagatorMomenta, Loops, Externals,}} \\
        &\text{\texttt{    NullMomenta, Replacement ];}}
    \end{aligned}
\end{align*}

Once a family is defined, \texttt{DivergentReduce} can be used to reduce the expansion coefficients of MIs:
\begin{align*}
    &\text{\texttt{DivergentReduce[ Family, "Level" -> n,}}\\
    &\text{\texttt{"Approach"->app ];}}
\end{align*}
The default value of the optional parameter \texttt{"Approach"} is \texttt{"DRR"}, to denote constructing finite FIs in high dimensions and then using DRR, and it can also be set as \texttt{"4d"} to denote constructing finite FIs in 4 dimensions directly. The optional parameter \texttt{"Level"} specifies the order to which MIs should be expanded, whose default value is $-1$. \texttt{DivergentReduce} will expand the MIs to $\mathcal{O}(\epsilon^n)$ order and reduce these expansion coefficents to MCs. After \texttt{DivergentReduce} completes, call \texttt{Family["DivMaster"]} to display the MCs.

Additionally, \textbf{FFI} alse provides methods to perform UV subtractions: 
\begin{align*}
    \begin{aligned}
        &\text{\texttt{BurnUV[ Family, M ];}} \\
        &\text{\texttt{UVCounterTerm[ Integral, Family ];}} 
    \end{aligned}
\end{align*}

More details of the use of the package can be found in the examples of the package \textbf{FFI}.

\sect{Summary and outlook}
\label{sect:summary}

In this article, we develop a systematic method to construct as many finite Feynman integrals (FIs) as possible, which can be used to reduce the $\epsilon$-expanded master integrals (MIs) to a minimal basis. There are two approaches to obtain infrared (IR) and collinear (CO) finite FIs: either by simply increasing the spacetime dimension of a given seed FI or by constructing appropriate numerators using the method described in Refs.~\cite{Gambuti:2023eqh, delaCruz:2024xsm}. We introduce an IR and CO regulator $\eta$ and employ the \texttt{asy2} package to verify finiteness in the IR and CO regions. For ultraviolet (UV) divergences, we propose a $\bar R$-operation method, an extension of the $R$-operation that ensures IR and CO safety in the subtraction terms.

Our method has been implemented in the open-source package \texttt{FFI}, which facilitates the discovery of relations among MIs in their $\epsilon$-expansion form.  
As demonstrated in the examples, the number of MCs is significantly smaller than the original set of MI expansion coefficients. Thus, the constraints introduced in our method effectively reduce the computational complexity of evaluating MIs.

\begin{acknowledgments}
    The work was supported in part by the National Natural Science Foundation of China (No. 12325503, No. 12357077) and the High-performance Computing Platform of Peking University.  
\end{acknowledgments}

\appendix
\section{UV divergence subtraction of two-loop sunrise}
\label{appendix:UV}
\begin{figure}[htbp]
    \centering
    \includegraphics[width=0.20\textwidth]{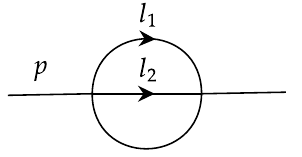}
    \caption{Two-loop sunrise.}
    \label{fig:sunrise}
\end{figure}
We use the two-loop sunrise as an example to demonstrate the calculation of UV counterterms. The propagators are:
\begin{equation}\label{eq:sunriseprop}
\begin{aligned}
 & D_1=l_1^2, \\
 & D_2=l_2^2, \\
 & D_3=(l_1+l_2-p)^2,
\end{aligned}
\end{equation}
and the ISPs are:
\begin{equation}\label{eq:sunriseISPs}
 D_4=p \cdot l_1,\quad D_5=p \cdot l_2,
\end{equation}
and $p^2 = s$.

Based on the approach in Sect. \ref{sect:IRandCO}, we present the following IR- and CO-finite integrand:
\begin{equation}\label{eq:sunriseIRfinite}
  G = \frac{D_4}{D_1^2 D_2 D_3},
\end{equation}
and our goal is to perform the UV subtraction and not to introduce new IR and CO divergences in the counterterms.

There are four UV regions in total:
\begin{align}\label{eq:sunriseUVregion}
    \begin{aligned}
      & r_1: 
      \{l_1+l_2=S,l_1 \rightarrow \infty\}, \\
      & r_2: 
      \{l_2=S,l_1 \rightarrow \infty\}, \\
      & r_3: 
      \{l_1=S,l_2 \rightarrow \infty\}, \\
      & r_4: 
      \{l_1,l_2 \rightarrow \infty\},
    \end{aligned}
\end{align}
then we have
\begin{align}
    \begin{aligned}
        &\mathcal{S}_1 = \{ l_1+l_2 \}, \\
        &\mathcal{S}_2 = \{ l_2 \}, \\
        &\mathcal{S}_3 = \{ l_1 \}, \\
        &\mathcal{S}_4 = \emptyset,
    \end{aligned}
\end{align}
and $\mathcal{S}_1,\mathcal{S}_2,\mathcal{S}_3 \supsetneq \mathcal{S}_4$, which means $r_1,r_2,r_3$ are sub-regions of $r_4$. Therefore, the UV subtraction formula in the case of the two-loop sunrise can be written as:
\begin{align}\label{eq:sunriseUVformula}
\begin{aligned}
  G_{\text{UVCT}} = & \ G-G_1-G_2-G_3-G_4 \\
  &+G_{14}+G_{24}+G_{34}.
\end{aligned}
\end{align}
Next we choose the $c_{ij}$'s in substitution~\eqref{eq:identitysubstitute} as follows:
\begin{align}
    \begin{aligned}
        &c_{11} = c_{22} = 1, \\
        &c_{12} = c_{21} = -\frac{1}{2}.
    \end{aligned}
\end{align}
Under the substitution, the $G$ is 
\begin{align}
    G = \frac{p\cdot l_1}{(\hat{l}_1^2 + m^2)^2(\hat{l}_2^2 + m^2)((\hat{l}_1+\hat{l}_2-p)^2+m^2)}.
\end{align}
Now we need to compute each term appearing in~\eqref{eq:sunriseUVformula}. Take $G_3$ as an example: In region $r_3$, $l_1=S, \hat{l}_1=\hat{S}, l_2=L, \hat{l}_2=\hat{L}$ and $G$ can be rewritten as
\begin{align}
    \begin{aligned}
         &G(L, S, \hat{L}, \hat{S}) \\
         =& \frac{p \cdot S}{( \hat{S}^2 +m^2)^2 (\hat{L}^2 + m^2)((\hat{S}+\hat{L}-p)^2+m^2)},
    \end{aligned}
\end{align}
which leads to the following expansion:
\begin{align}
    G(\Lambda L, S, \Lambda\hat{L}, \hat{S}) = \frac{1}{\Lambda^4} \frac{p\cdot S}{(\hat{S}^2+m^2)^2(\hat{L}^2)^2} + \mathcal{O}(\frac{1}{\Lambda^5}),
\end{align}
and therefore
\begin{align}
    G_3 = \frac{p\cdot S}{(\hat{S}^2+m^2)^2(\hat{L}^2)^2} = \frac{ l_1 \cdot p}{(l_1^2)^2 (l_2^2-m^2)^2}.
\end{align}
Similarly, we can obtain the expression for $G_1, G_2, G_4$ and $G_{14},G_{24},G_{34}$:
\begin{align}\label{eq:sunriseG1-4}
    \begin{aligned}
        G_1&=0, \quad G_2=0, \\
        G_4&=\frac{l_1 \cdot p}{(l_1^2-m^2)^2 (l_2^2-m^2) ((l_1+l_2)^2-m^2)} \\
        &+\frac{2(l_1 \cdot p)^2}{(l_1^2-m^2)^2 (l_2^2-m^2) ((l_1+l_2)^2-m^2)^2} \\
        &+\frac{2(l_1 \cdot p)(l_2 \cdot p)}{(l_1^2-m^2)^2 (l_2^2-m^2) ((l_1+l_2)^2-m^2)^2}, \\
        G_{14} &=0,\quad G_{24}=0, \\
        G_{34} &= (G_3)_4 =\frac{l_1 \cdot p}{(l_1^2-m^2)^2 (l_2^2-m^2)^2}.
    \end{aligned}
\end{align}

\section{DRR and IBP in 2-loop double box}
\label{appendix:dbdrr}
By only retaining the coefficients of the MIs in the topsector, we get:
\begin{align}
    &\begin{aligned}
        &I^{(D)}(1,1,1,1,1,1,2,0,0) \\
        =& -\frac{(-5+D)(-14+3D)}{(-6+D)t}M_1^{(D)} \\
        &- \frac{2(-5+D)(-4+D)}{(-6+D)st}M_2^{(D)} \\
        &+ \text{(subsectors)},
    \end{aligned} \\
    &\begin{aligned}
        M_1^{(D+2)} =& \ \frac{s^2 ((-10 + 3 D) s +  (-6 + 2D) t)}{8 (-3 + D)^3 (s + t)}M_1^{(D)} \\
        &- \frac{s^2 ((-30 + 9 D) s + (- 26 + 8 D)t)}{8 (-3 + D)^3 t (s + t)}M_2^{(D)} \\
        &+ \text{(subsectors)},
    \end{aligned} \\
    &\begin{aligned}
        M_2^{(D+2)}  =& - \frac{s^3 ((80 - 54 D + 9 D^2) s + 8 (-3 + D)^2 t)}{16 (-3 + D)^3 (-2 + D) (s + t)}M_1^{(D)} \\
        &+ (s^2 (3 (80 - 54 D + 9 D^2) s^2 \\
        &\quad+ 2 (140 - 92 D + 15 D^2) s t + 4 (12 - 7 D + D^2) t^2)) \\
        &\quad M_2^{(D)} / (16 (-3 + D)^3 (-2 + D) t(s + t)) \\
        &+ \text{(subsectors)}.
    \end{aligned}
\end{align}


\appendix

\providecommand{\href}[2]{#2}\begingroup\raggedright\endgroup

\end{document}